\documentclass[aip,amsmath,amssymb,reprint,]{revtex4-1}

\usepackage{graphicx}
\usepackage{dcolumn}
\usepackage{bm}
\usepackage[utf8]{inputenc}
\usepackage[T1]{fontenc}
\usepackage{mathptmx}
\usepackage{etoolbox}
\usepackage{textcomp}
\usepackage{float}
\usepackage{ulem}
\usepackage{color}
\usepackage{hyperref}
\newcounter{aqctr}
\newenvironment{author-query}
{\refstepcounter{aqctr}\par\vspace{\baselineskip}\noindent
\color{red}\textbf{Author Query/Comment AQ \arabic{aqctr}.}}
{\par\vspace{\baselineskip}\normalcolor}

\makeatletter
\def\@email#1#2{
 \endgroup
 \patchcmd{\titleblock@produce}
  {\frontmatter@RRAPformat} {\frontmatter@RRAPformat{\produce@RRAP{*#1\href{mailto:#2}{#2}}}\frontmatter@RRAPformat}
  {}{}
}
\makeatother
\begin{document}

\preprint{AIP/123-QED}

\title[Physics of Fluids]{The effects of cavitation position on the velocity of a laser-induced microjet extracted using explainable artificial intelligence}

\author{Daichi Igarashi}
\author{Jingzu Yee}
\author{Yuto Yokoyama}
\author{Hiroaki Kusuno}
\author{Yoshiyuki Tagawa}
 \altaffiliation{Authors to whom correspondence should be addressed: tagawayo@cc.tuat.ac.jp; Also at the Institute of Global Innovation Research, Tokyo University of Agriculture and Technology, Koganei, Tokyo 184-8588, Japan}

\affiliation{
Department of Mechanical Systems Engineering,
Tokyo University of Agriculture and Technology, 
Koganei, Tokyo 184-8588, Japan}

\begin{abstract}

The control of the velocity of a high-speed laser-induced microjet is crucial in applications such as needle-free injection.
Previous studies have indicated that the jet velocity is heavily influenced by the volumes of secondary cavitation bubbles generated through laser absorption.
However, there has been a lack of investigation of the relationship between the positions of cavitation bubbles and the jet velocity.
In this study, we investigate the effects of cavitation bubbles on the jet velocity of laser-induced microjets extracted using explainable artificial intelligence (XAI).
An XAI is used to classify the jet velocity from images of cavitation bubbles and to extract features from the images through visualization of the classification process.
For this purpose, we run 1000 experiments and collect the corresponding images.
The XAI model, which is a feedforward neural network (FNN), is trained to classify the jet velocity from the images of cavitation bubbles.
After achieving a high classification accuracy, we analyze the classification process of the FNN.
The predictions of the FNN, when considering the cavitation positions, show a higher correlation with the jet velocity than the results considering only cavitation volumes.
Further investigation suggested that cavitation that occurs closer to the laser focus position has a higher acceleration effect.
These results suggest that the velocity of a high-speed microjet is also affected by the cavitation position.

\end{abstract}

\maketitle

\section{Introduction}\label{sec:Intro}

Currently, injection needles are one of the most common devices used for drug delivery.
However, the use of needles for drug delivery can cause problems such as trypanophobia \cite{nir2003fear,mclenon2019fear,deacon2006fear} and the spread of infectious diseases.\cite{kermode2004unsafe,weiss2004social,resources2003incorporating}
Therefore, a needle-free injector, which injects liquid as a high-speed microjet into the body, has been developed in recent years. \cite{mitragotri2005immunization,stachowiak2009dynamic,schoppink2022jet,daly2020needle}
Commercialized needle-free injectors, which commonly use a diffused jet, are reported to cause significant pain during injection.\cite{harding2002comparison, giudice2006needle,mitragotri2006current,arora2007needle}
To solve this problem, a laser-induced microjet with a focused tip is expected to be useful because of the lower stress induced at the injection site and the higher efficiency of liquid transportation.\cite{tagawa2012highly, tagawa2013needle, miyazaki2021dynamic,kiyama2019visualization}
To develop an efficacious needle-free injector using a focused microjet, it is necessary to control the jet velocity.

Many studies have been carried out with the aim of understanding the mechanism that affects jet velocity.\cite{franco2021effect,van2010breakup,scroggs1996experimental,choo2007effect,thoroddsen2009spray, kamamoto2021drop}
Several of these studies have reported that the total volumes of secondary cavitation bubbles in the liquid affects the jet velocity, and have proposed a linear model to explain the relationship.\cite{kiyama2016effects,hayasaka2017,daily2014catastrophic,ishikawa2022numerical}
According to a study by Kiyama \textit{et al.},\cite{kiyama2016effects} this process can be explained as follows.
When a laser is absorbed by a liquid in a microtube, compression waves are generated, which accelerate the jet.
These compression waves are reflected at the gas-liquid interface as expansion waves, which decelerate the jet.
When cavitation bubbles occur at the positions where the pressure in the liquid falls below the saturation vapor pressure, the cavitation bubbles inhibit the propagation of the expansion waves.
Therefore, the greater the amount of cavitation, the more easily the propagation of the expansion wave is inhibited, and the faster the jet velocity.
However, plots in the same paper of the jet velocity versus the total volumes of cavitation bubbles still showed a non-negligible amount of variance.
After reviewing these previous experimental data, we have found that multiple cavitation bubbles appear at different positions.
Therefore, we argue that the cavitation bubbles' positions, as well as the total volumes of the cavitation bubbles, may affect the jet velocity.

To investigate the influence of both the volumes and positions of cavitation bubbles on the jet velocity, we propose the use of artificial intelligence (AI), which has recently been the focus of a great deal of attention from the fluid mechanics community,\cite{brunton2020machine, colvert2018classifying, ling2016reynolds, kutz2017deep} even in the field of cavitation bubbles.\cite{folden2023classification}
Feature extraction has been studied by fluid researchers in various fields, using methods such as dynamic mode decomposition (DMD),\cite{williams2015data,jovanovic2014sparsity} principal component analysis (PCA),\cite{abdi2010principal} and proper orthogonal decomposition (POD).\cite{kerschen2005method}
In addition, many fluid studies have been performed using convolutional neural networks (CNNs) for problems dealing with images.\cite{vennemann2020dynamic,sekar2019fast,liu2020deep,cai2019dense,jin2018prediction}
However, for most AI approaches, the underlying reasoning that leads to a specific decision is unknown or not properly understood.
In contrast, explainable AI (XAI),\cite{arrieta2020explainable, adadi2018peeking, gunning2017explainable, gunning2019xai, yee2023prediction} which does not have the problems of explainability and interpretability, represents a powerful tool that can aid in the understanding of physical phenomena.
Therefore, the objective of this study is to clarify the effects of bubble volume and position on jet velocity through feature extraction using a large amount of experimental data and XAI.
In a study by Yee \textit{et al.},\cite{yee2022image} weight visualization with XAI was performed to extract the features that differentiate splashing and non-splashing drops when a droplet impacts a solid surface.
We adopt the same method here to extract the positions of the cavitation bubbles that influence the jet velocity.

The structure of this paper is as follows.
In Sec.~\ref{sec:experiment}, the experimental setup and the collection of experimental data are described.
We also describe the relationship between the amount of cavitation bubbles and the jet velocity here.
Next, in Sec.~\ref{sec:ML_method}, the XAI implementation and the methods of feature extraction are explained.
We also show results from the training and testing of the XAI implementation.
In Sec.~\ref{sec:RanD}, we explain the prediction and feature extraction results of the FNN and discuss the results.
We analyze the effects of the positions of the cavitation bubbles on the jet velocity.
Finally, the conclusions and outlook of this study are presented in Sec.~\ref{sec:conclusion}.

\section{Experiments}\label{sec:experiment}

This section describes the methodology of the experiments and the relation between the amount of secondary cavitation bubbles and the jet velocity.

\subsection{Experimental setup}\label{subsec:exp_setup}

\begin{figure}[t]
\centering
\includegraphics[width=1.0\columnwidth]{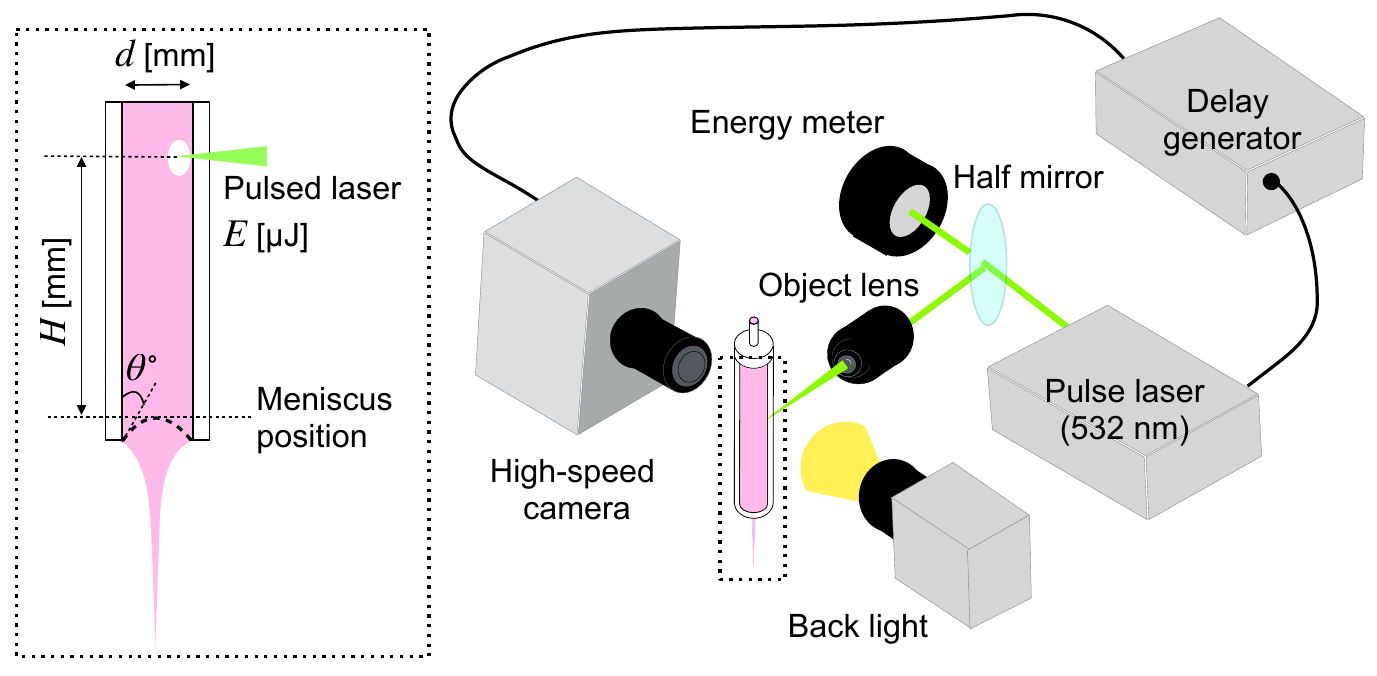}
\caption{\label{fig:exp_setup}
A schematic diagram of the experimental setup for the laser-induced microjet.}
\end{figure}

The experimental setup is shown in Fig.~\ref{fig:exp_setup}. 
It is based on a study by Tagawa \textit{et al.} \cite{tagawa2012highly}
A green pulsed laser generated by a laser generator (Nd: YAG laser Nano SPVI, Litron Laser Co., Ltd., wavelength: 532 nm) was passed through a mirror, a half mirror (Dielectric plate half-mirror, OptoSigma Co., transmission: 50\%), and an objective lens (MPLn10x, Olympus Co. Ltd., Japan, magnification: 10x, N.A. value: 0.25). 
We used magenta ink (GI-30M, Canon), which has high green laser absorbance, as the working liquid.
The laser energy was measured with an energy meter (ES111C, Thorlabs, energy range: 10 \textmu J--150 mJ).
The image sequences of the laser-induced microjet were filmed with a high-speed camera (FASTCAM SA-X, Photron Co., Ltd.).
The temporal resolution of the camera was 100\,000 f.p.s. 
The lens was attached with a high-pass filter (SCF-50S-56O, Sigma Koki Co., transmission limit wavelength: 560 nm) to shade the 532 nm laser light. 
A light source (SLG-150V-CW-MN, REVOX Inc.) was used as the backlight for the high-speed camera.
The laser generator and the high-speed camera were synchronized using a delay generator (Model 575 Pulse/Delay generator, BNC Co.). 
In this study, we acquired 1000 image sequences of the laser-induced microjet.

According to the study by Tagawa \textit{et al.},\cite{tagawa2012highly} the parameters that affect the jet velocity are the laser energy $E$ [\textmu J], the inner diameter of the microtube $d$ [mm], the contact angle between the meniscus interface and the inner wall $\theta$ [$^\circ$], and the distance from the meniscus interface to the laser irradiation position $H$ [mm], as shown in Fig.~\ref{fig:exp_setup}.
In this study, the ranges of these parameters are $E=$ 500--600 \textmu J, $d=$ 0.5 mm, $\theta=$ 40.3--59.4$^\circ$, and $H=$ 1.84--2.09 mm, respectively.
In addition, the same tube was used throughout the experiments.

\subsection{Experimental observations}\label{subsec:microjet}

\begin{figure}[t]
\centering
\includegraphics[width=1.0\columnwidth]{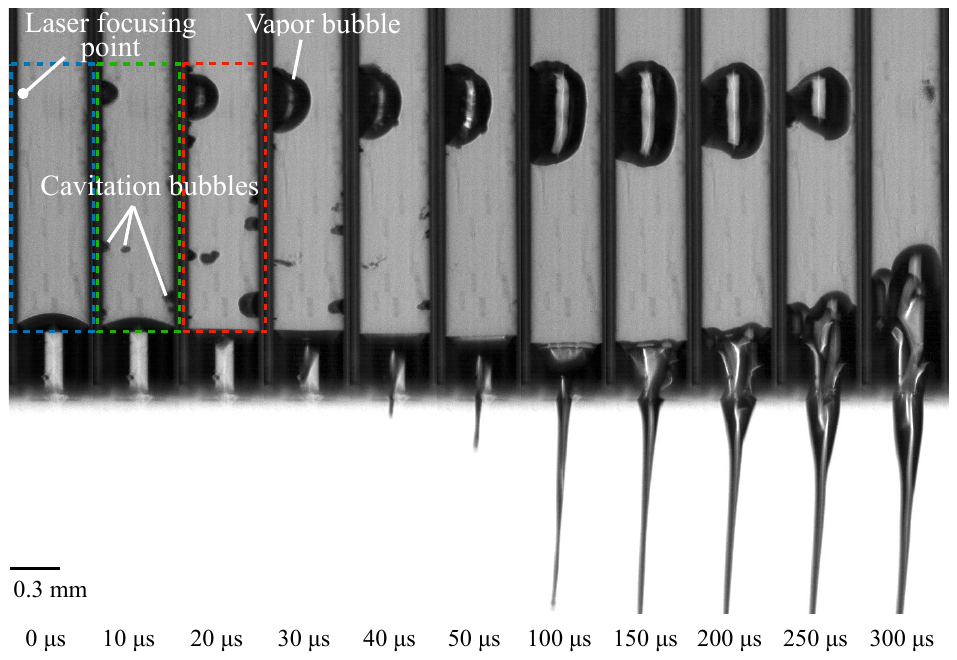}
\caption{\label{fig:microjet}
An image sequence of the laser-induced microjet is obtained from the experiment. The vapor bubble is observed at $t=$ 10--250 \textmu s, the cavitation bubbles are observed at $t=$ 10--40 \textmu s, and the microjet is observed at $t=$ 20--300 \textmu s.}
\end{figure}

Fig.~\ref{fig:microjet} shows an example of the image sequence taken from the experiments described above. 
We set the time of the frame immediately before the laser illumination as $t=0$ \textmu s.
A laser-induced microjet was generated after the laser was focused on the liquid ($t=$ 20--300 \textmu s). 
During this process, we observed an expanding bubble at the laser focusing point ($t=$ 10--300 \textmu s). 
This bubble, known as a vapor bubble, was created when the liquid was vaporized by the heat of the laser. \cite{akhatov2001collapse}
After a certain period of time, the vapor bubble began to shrink ($t\approx200$ \textmu s) and finally collapsed ($t\approx300$ \textmu s).

In addition to the vapor bubble, we also observed small secondary bubbles in the images at $t=$ 10--40 \textmu s in Fig.~\ref{fig:microjet}.
As shown in the figure, these secondary bubbles generated at $t\approx$ 10 \textmu s began to shrink at $t\approx$ 30 \textmu s and eventually disappeared at $t\approx$ 50 \textmu s.
Such bubbles, known as cavitation bubbles, are caused by a rapid phase change from liquid to gas due to the decrease in the pressure of the liquid.\cite{caupin2006cavitation}
Previous studies have indicated that the jet velocity is related to these cavitation bubbles. \cite{kiyama2016effects,hayasaka2017,daily2014catastrophic,ishikawa2022numerical}
Therefore, we investigated the images in the frame immediately before the laser illumination ($t=0$ \textmu s), the frame immediately after the laser illumination ($t=10$ \textmu s), and the succeeding frame ($t=20$ \textmu s).
At $t=0$ \textmu s, there were no cavitation bubbles. 
In the frames at $t=10$ \textmu s, cavitation bubbles started to form.
In the frame at $t=20$ \textmu s, the cavitation bubbles reached their maximum sizes in most cases, as shown in Fig.~\ref{fig:microjet}.

\subsection{The effects of the laser energy and amount of cavitation on the jet velocity}\label{subsec:VandS_result}

\begin{figure}[t]
\centering
\includegraphics[width=0.8\columnwidth]{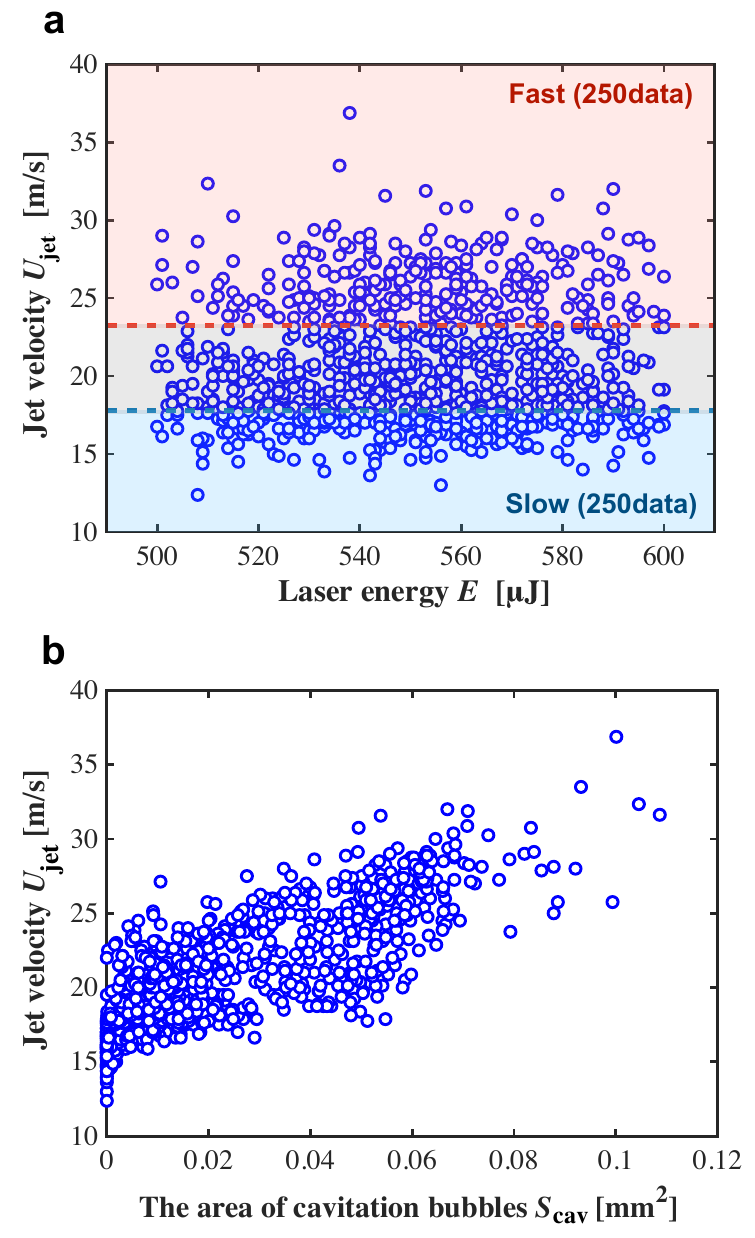}
\caption{\label{fig:exp_result}
(a) The relationship between the jet velocity $U_\mathrm{\mathrm{jet}}$ and laser energy $E$ in the experiment, where the blue and red areas represent the data of the fastest 25\% and the slowest 25\% of the jet velocity, respectively. (b) The relationship between the jet velocity $U_\mathrm{\mathrm{jet}}$ and the total projected area of the cavitation bubbles $S_\mathrm{\mathrm{cav}}$ at $t = 20$ \textmu s.}
\end{figure}

The relationship between the jet velocity $U_\mathrm{jet}$ and the laser energy $E$ for the 1000 data is shown in Fig.~\ref{fig:exp_result}(a).
As shown in the figure, there is significant variation in the jet velocity for a fixed laser energy.
By comparing images with the same $E$ but different $U_\textrm{jet}$, we found some images with large amounts of cavitation bubbles, some with small amounts of cavitation bubbles, and others with no cavitation bubbles.
Kiyama \textit{et al.}\cite{kiyama2016effects} pointed out that the jet velocity increases with the total volumes of cavitation bubbles in the liquid.
Therefore, we investigated the relationship between the jet velocity $U_\mathrm{\mathrm{jet}}$ and the total projected area of cavitation bubbles $S_\mathrm{\mathrm{cav}}$ of the experimental data when the cavitation bubbles reached their maximum sizes, i.e., $t=20$ \textmu s.
The results are shown in Fig.~\ref{fig:exp_result}(b).
We calculated the cavitation area $S_\mathrm{\mathrm{cav}}$ from the number of pixels occupied by cavitation bubbles in the binarized input images.
It was observed that the two roughly correlate with each other, with the correlation coefficient of the jet velocity and the total area of cavitation bubbles being 0.817.
However, there is still a non-negligible variance.
A possible contributing factor to this issue could be the lack of consideration of the positions of the cavitation bubbles.

\section{Explainable artificial intelligence (XAI): feedforward neural network (FNN)}\label{sec:ML_method}

To investigate the effects of the cavitation position on the jet velocity, we used explainable artificial intelligence (XAI) to classify the jet velocity from images with cavitation bubbles and extract the features of the images through visualization of the classification process.
In this section, we describe the implementation of the XAI and the visualization of the classification process.

\begin{figure*}[t]
\centering
\includegraphics[width=0.8\textwidth]{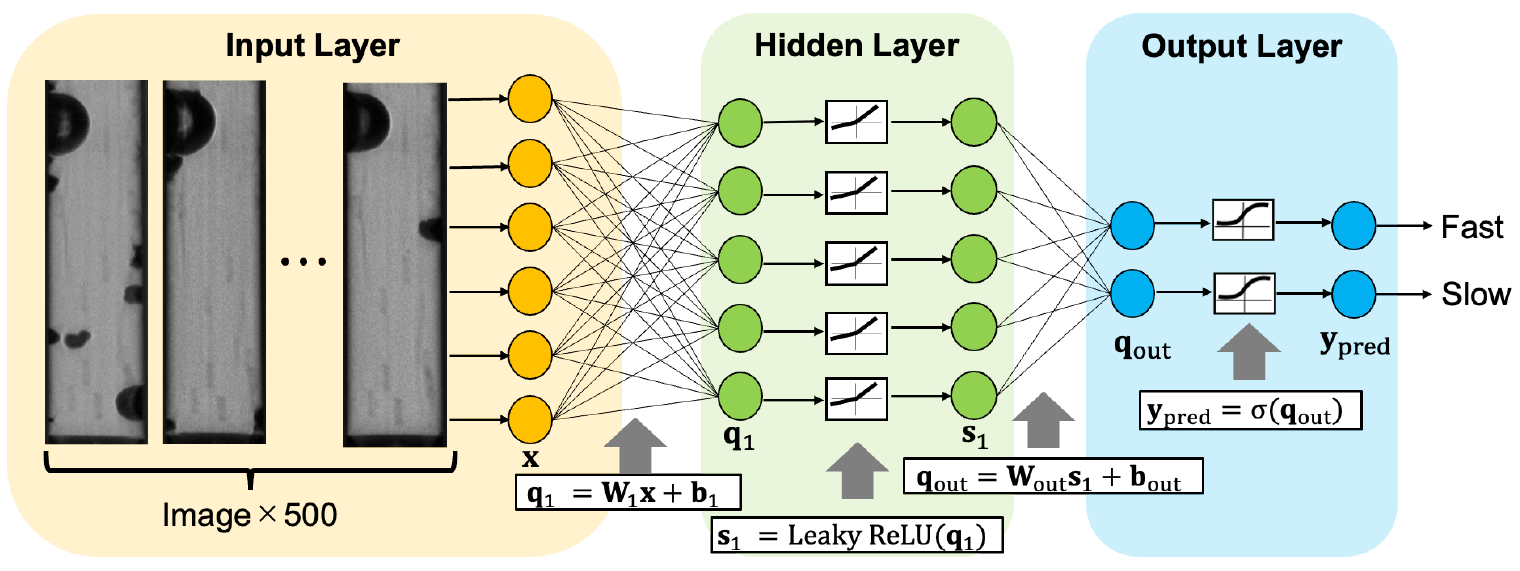}
\caption{\label{fig:FNN}
The architecture of the feedforward neural network for classifying the ``Fast'' and ``Slow'' jet velocity classes in this study.}
\end{figure*}

\subsection{Implementation of the FNN}\label{subsec:FNN}

For the XAI, we used the architecture of a feedforward neural network(FNN),\cite{rosenblatt1958perceptron} as shown in Fig.~\ref{fig:FNN}.
An FNN is a simple neural network in which the decision-making process can be explained through visualization of the classification process.\cite{yee2022image}
According to a study by Hornik and Stinchcombe,\cite{hornik1989multilayer} an FNN with three or more layers can approximate a continuous function with arbitrary precision (universal approximation theorem).
Therefore, we used a three-layered FNN.
In addition, we adopted the method of Yee \textit{et al.}\cite{yee2022image} to address the problem as a binary classification problem to predict whether the jet velocity was ``Fast'' or ``Slow'', and visualized the image features of each classification.

The captured images were preprocessed before being input into the FNN.
First, we cropped the image of all experimental data to focus on the phenomena in the microtube.
The cropping range for each image was 391$\times$111 pix in the area enclosed by the blue, green, and red dashed boxes in Fig. \ref{fig:microjet}.
Next, we normalized the brightness values of the images.
To feed an image into the FNN, each two-dimensional matrix of the input image was flattened into a one-dimensional vector.
The number of neurons in the input layer equals the total number of pixels in an input image.
The input vectors of the hidden layer are defined as $\mathbf{q_\mathrm{1}}$:
\begin{equation} 
\label{eq:xtoq1}
\mathbf{q_\mathrm{1}=W_\mathrm{1}x+b_\mathrm{1}}.
\end{equation}
\noindent Here, $\mathbf{W_\mathrm{1}}$ and $\mathbf{b_\mathrm{1}}$, respectively, represent the weight matrix and bias vector that connect the input and hidden layers.
In addition, $\mathbf{x}$ denotes the flattened one-dimensional vector of an input image. 

We set the number of neurons in the hidden layer to five, which is the optimal number obtained through parameter tuning.
In the hidden layer, we used the leaky rectified linear unit (Leaky ReLU) function \cite{maas2013rectifier} as the activation function.
By representing the output vectors of the hidden layer with $\mathbf{s_\mathrm{1}}$, the equation in the hidden layer could be expressed as
\begin{equation} 
\label{eq:hidden}
\mathbf{s_\mathrm{1}} = \mathrm{Leaky\,ReLU}(\mathbf{q_\mathrm{1}})=
\begin{cases}
\mathbf{q_\mathrm{1}} & \textrm{if}\;\mathbf{q_\mathrm{1}} \geq 0\\
\alpha \mathbf{q_\mathrm{1}} & \textrm{if}\;\mathbf{q_\mathrm{1}} < 0,\\
\end{cases}
\end{equation}
\noindent where $\alpha$ is set to 0.01, which is the optimal value as determined by hyperparameter tuning.
 
By defining the input vectors of the output layer as $\mathbf{q_\mathrm{out}}$, the equation that connects the hidden and output layers can be expressed as
\begin{equation} 
\label{eq:s1toqout}
\mathbf{q_\mathrm{out}=W_\mathrm{out}s_\mathrm{1}+b_\mathrm{out}}.
\end{equation}
\noindent Here, $\mathbf{W_\mathrm{out}}$ and $\mathbf{b_\mathrm{out}}$ represent the weight matrix and bias vector that connect the hidden and output layers.

For the classification problem, out of the 1000 data, we labeled the 250 with the highest velocities ($U_{\rm{jet}} > 23.8$ m/s) as ``Fast'' and the 250 with the lowest velocities ($U_{\rm{jet}} < 17.6$ m/s) as ``Slow'', as shown in Fig.~\ref{fig:exp_result}(a).
Note that the blue and red areas in the figure represent the data of the fastest 25\% and the slowest 25\% of the jet velocity, respectively.
Thus, the number of neurons in the output layer was two, i.e., the number of classes.

In the output layer, we used the sigmoid function as the activation function. 
By defining the output vectors of the output layer as $\mathbf{y_\mathrm{pred}}$, the equation in the output layer could be expressed as
\begin{equation} 
\label{eq:output}
\mathbf{y}_\mathrm{pred} = \mathrm{\sigma} (\mathbf{q_\mathrm{out}}) = \frac{1} {1+\exp(-\mathbf{q_\mathrm{out}})}.
\end{equation}
\noindent Here, we define the two elements of the $\mathbf{y}_\mathrm{pred}$ vector, which give the probabilities of an image being predicted to be ``Fast'' and ``Slow'', as $y_\mathrm{pred1}$ and $y_\mathrm{pred2}$, respectively.
A prediction of ``Fast'' is made when $y_\mathrm{pred1}>y_\mathrm{pred2}$ and a prediction of ``Slow'' is made when $y_\mathrm{pred1}<y_\mathrm{pred2}$.

\subsection{Training and testing}\label{subsec:tandt}

\begin{figure}[t]
\centering
\includegraphics[width=1.0\columnwidth]{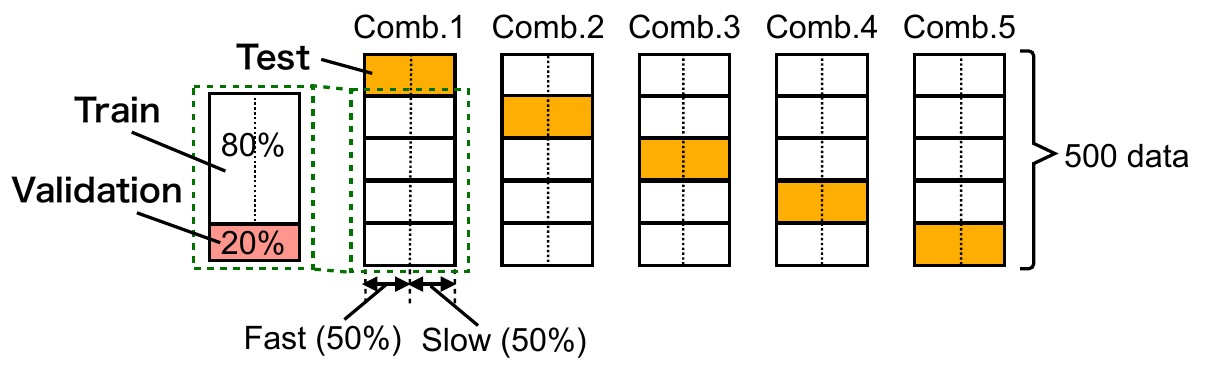}
\caption{\label{fig:cross_val}
An illustration of the cross-validation: the combinations of training, validation, and test data used in this study (where comb. means combination).}
\end{figure}

We evaluated the generalization performance of the FNN using cross-validation.\cite{geron2019hands}
As explained in Sec.~\ref{subsec:FNN}, we used 250 ``Fast'' data and 250 ``Slow'' data as the training data for the classification problem.
Note that the remaining 500 data were used as the interpolation test data in Sec.~\ref{sec:RanD}.
As shown in Fig.~\ref{fig:cross_val}, we divided the 500 training data into five parts of 100.
For a given data combination, one of the five parts was used for testing, while the remaining four parts were used for training and validation.
Therefore, there were five data combinations in total.
We set the number of ``Fast'' and ``Slow'' data to be equal in both the training and testing datasets for each combination.

\begin{figure}[t]
\centering
\includegraphics[width=0.8\columnwidth]{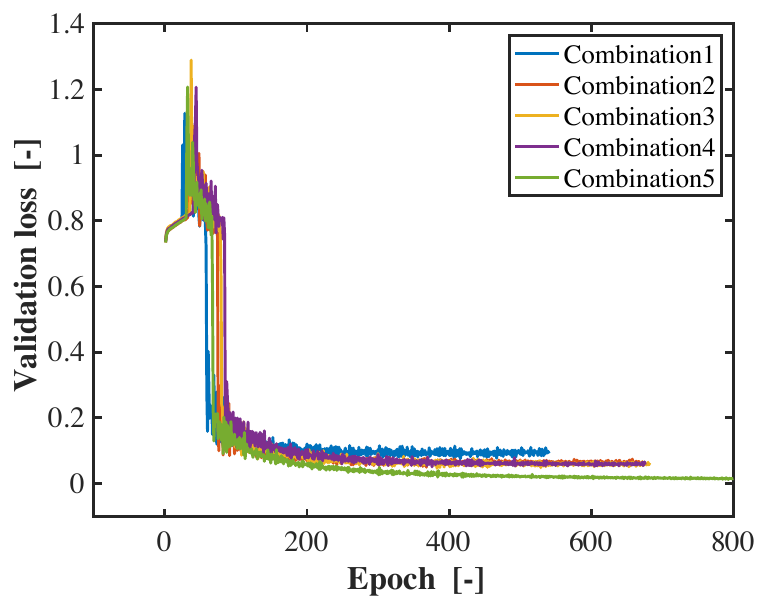}
\caption{\label{fig:tandv_result}
An evaluation of the validation losses for every epoch of the training using each combination. Note that Combination 5 continued to train until about 2500 epochs.}
\end{figure}

We trained the neural network described in Sec.~\ref{subsec:FNN} with momentum stochastic gradient descent (SGD)\cite{li2014efficient} as the optimizer.
We performed mini-batch training with the number of data in each batch set to 32.
The validation data used was a randomly selected 20\% of all data excluding test data.
The evaluation of the validation losses for each epoch is shown in Fig.~\ref{fig:tandv_result}.
Here, the number of epochs represents how many times an entire training dataset is fed to the FNN.
The figure shows that the error decreases with each epoch for all combinations.
Therefore, the training validated the generalization performance.
Note that in order to avoid overfitting the training data, we used early stopping,\cite{prechelt2002early} a method that terminates the training process when the error increases over that of the previous epoch.

After training, we investigated the test accuracy of the classification for the jet velocity from the trained FNN using images at different times.
Tables~\ref{tab:test_result_0}, \ref{tab:test_result_10}, and \ref{tab:test_result_20} show the accuracy when using images at $t=0$ \textmu s (the blue dashed box in Fig.~\ref{fig:microjet}), $t=10$ \textmu s (the green dashed box in Fig.~\ref{fig:microjet}), and $t=20$ \textmu s (the red dashed box in Fig.~\ref{fig:microjet}), respectively.
Each table shows the test accuracy of the ``Fast'', ``Slow'', and ``Total'' predictions of each combination and the average accuracy of all combinations.

\begin{table}[t]
\centering
\caption{\label{tab:test_result_0} 
The test accuracy of the ``Fast'' and ``Slow'' predictions in each combination for $t=0$ \textmu s.}
\begin{tabular}{ccccccccc}
\hline\hline
Combination & Fast & Slow & Total\\
\hline
1& 50/50 (100\%) & 0/50 (0\%) & 50/100 (50\%)\\ 
2& 50/50 (100\%) & 0/50 (0\%) & 50/100 (50\%)\\ 
3& 50/50 (100\%) & 0/50 (0\%) & 50/100 (50\%)\\ 
4& 47/50 (94\%) & 11/50 (22\%) & 58/100 (58\%)\\ 
5& 47/50 (94\%) & 21/50 (42\%) & 68/100 (68\%)\\
\hline
Average & 97.6\% & 12.8\% & 55.2\%\\ 
\hline\hline
\end{tabular}
\end{table}

At $t=0$ \textmu s, Table~\ref{tab:test_result_0} shows that most combinations have a total accuracy of 50\%.
This is the minimum prediction accuracy because there are only two classification classes and the number of ``Fast'' and ``Slow'' test data is equal.
Therefore, this accuracy can be obtained just by classifying all the test data as one of the two classes.
From above, with regard to the accuracy using the images at $t=0$ \textmu s, the FNN predicted almost all data as ``Fast'', which means the jet velocity was not accurately predicted.
This is because there is not much difference between the images at $t=0$ \textmu s.

Next, we compared the test results at $t=10$ \textmu s (Table~\ref{tab:test_result_10}) and $t=20$ \textmu s (Table~\ref{tab:test_result_20}).
As shown in Table~\ref{tab:test_result_10}, at $t=10$ \textmu s, the accuracy of the ``Fast'' and ``Slow'' cases is higher than 70.0\% for all five combinations.
Although it is higher than that at $t=0$ \textmu s, there are still inaccurate predictions in some cases.
However, as shown in Table~\ref{tab:test_result_20}, these wrongly predicted data are correctly predicted when images at $t=20$ \textmu s are used for training.
Therefore, we inspected the ``Fast'' data that were wrongly predicted using images at $t=10$ \textmu s but correctly predicted using images at $t=20$ \textmu s.
The investigation shows that for most of these data, cavitation bubbles appeared in the images at $t=20$ \textmu s but did not in the images at $t=10$ \textmu s.
This suggests that the presence of cavitation bubbles affects the test accuracy.
Finally, the averaged total accuracy was 99.2\%, indicating that the FNN extracted a straightforward feature that affects the jet velocity.
Note that the increase in the accuracy of the ``Fast'' data at $t=20$ \textmu s improves the accuracy of the ``Slow'' data because of the nature of a binary classification problem.
In this study, to understand the effects of the cavitation bubbles, we investigate the images at $t=20$ \textmu s when the volumes of the cavitation bubbles and the averaged total accuracy are maximal among $t=0$ \textmu s to $t=150$ \textmu s.

\begin{table}[t]
\centering
\caption{\label{tab:test_result_10} 
The test accuracy of the ``Fast'' and ``Slow'' predictions in each combination for $t=10$ \textmu s.}
\begin{tabular}{ccccccccc}
\hline\hline
Combination & Fast & Slow & Total\\
\hline
1& 38/50 (76\%) & 47/50 (94\%) & 85/100 (85\%)\\ 
2& 49/50 (98\%) & 46/50 (92\%) & 95/100 (95\%)\\ 
3& 50/50 (100\%) & 41/50 (82\%) & 91/100 (91\%)\\ 
4& 50/50 (100\%) & 42/50 (84\%) & 92/100 (92\%)\\ 
5& 50/50 (100\%) & 35/50 (70\%) & 85/100 (85\%)\\ 
\hline
Average & 94.8\% & 84.4\% & 89.6\%\\ 
\hline\hline
\end{tabular}
\end{table}

\begin{table}[t]
\centering
\caption{\label{tab:test_result_20} 
The test accuracy of the ``Fast'' and ``Slow'' predictions in each combination for $t=20$ \textmu s.}
\begin{tabular}{ccccccccc}
\hline\hline
Combination & Fast & Slow & Total\\
\hline
1& 49/50 (98\%) & 50/50 (100\%) & 99/100 (99\%)\\ 
2& 50/50 (100\%) & 50/50 (100\%) & 100/100 (100\%)\\
3& 50/50 (100\%) & 50/50 (100\%) & 100/100 (100\%)\\
4& 50/50 (100\%) & 49/50 (99\%) & 99/100 (99\%)\\ 
5& 48/50 (96\%) & 50/50 (100\%) & 98/100 (98\%)\\ 
\hline
Average & 98.8\% & 99.6\% & 99.2\%\\ 
\hline\hline
\end{tabular}
\end{table}

\begin{figure*}[!t]
\centering
\includegraphics[width=0.8\textwidth]{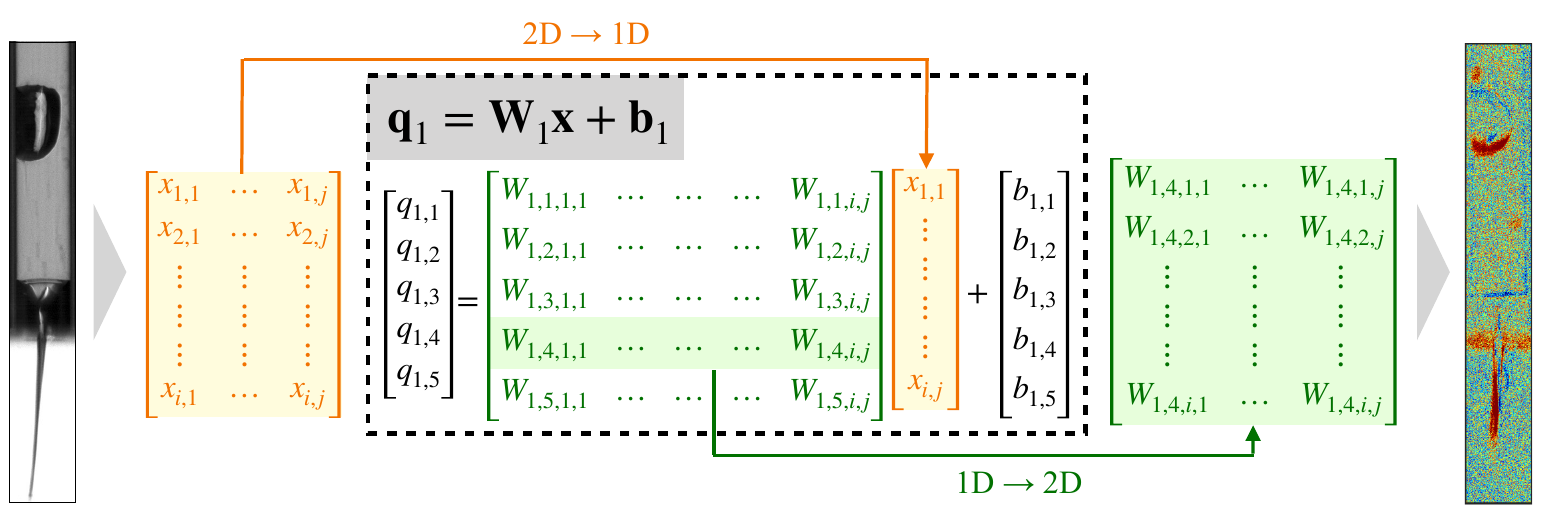}
\caption{\label{fig:weight_vis}
Weight visualization is performed by reshaping each row vector of the weight matrix $\bf{W}_1$ into the shape of the input images: the orange arrow represents a two-dimensional to one-dimensional reshaping and the green arrow represents a one-dimensional to two-dimensional reshaping.}
\end{figure*}

\subsection{Weight visualization}\label{subsec:vis_pro}
In this subsection, we explain the method for weight visualization and the meaning of positive and negative values in the visualized weight.

We describe the visualization method of the weight matrices $\mathbf{W_\mathrm{1}}$ for investigating the extracted features.\cite{yee2022image}
Since weights represent the importance of an input signal, visualization of the weight matrices enables us to investigate the important features of the input images. 

An illustration of the weight visualization method used in this study is shown in Fig.~\ref{fig:weight_vis}.
As shown by the orange arrow in the figure, the two-dimensional matrix of the input image is flattened into a one-dimensional vector and connected to neurons in the input layer. 
This one-dimensional matrix $\mathbf{x}$ is input into Eq.~\ref{eq:xtoq1}.

\begin{figure}[t]
\includegraphics[width=1.0\columnwidth]{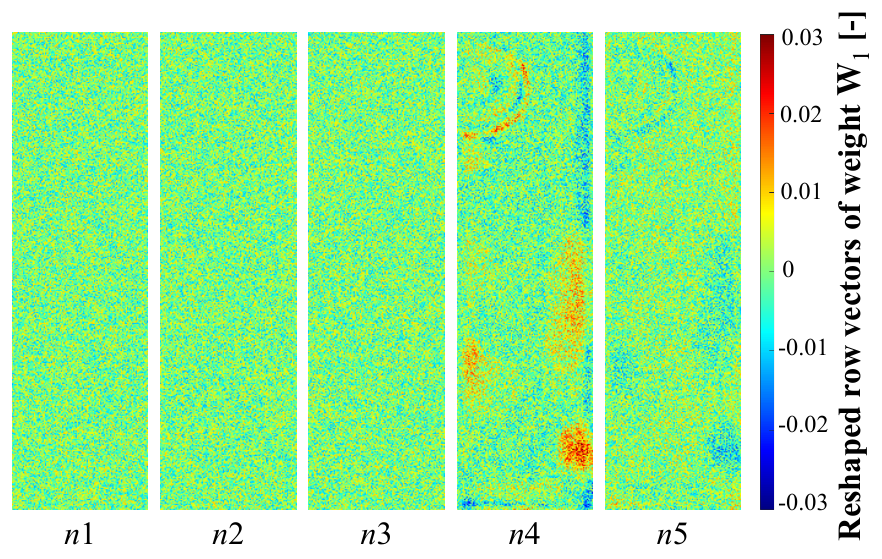}
\caption{\label{fig:qout_s1} 
Weight visualization for the five neurons ($n1$--$n5$).}
\end{figure}

As explained in Sec.~\ref{subsec:FNN}, we set the number of neurons in the hidden layer to five in this study, so the number of rows in the weight matrix $\mathbf{W_\mathrm{1}}$ is five.
Note that we refer to the five neurons as $n1$, $n2$, $n3$, $n4$, and $n5$.
As shown by the green arrow in the figure, the weight vector that corresponds to each of these neurons is reshaped into a two-dimensional matrix, which has the same size as an input image and is visualized as a colormap.
The colormaps of the reshaped weight vectors of the neurons trained using Combination 5 are shown in Fig.~\ref{fig:qout_s1}.
Note that we used Combination 5 for the explanation because the extracted features in this case are the clearest among all combinations.
The results are similar for all combinations.

In Fig.~\ref{fig:qout_s1}, red, green, and blue indicate positive, zero, and negative values, respectively.
Although there are almost all zero values in $n1$, $n2$, and $n3$, the distributions of the positive and negative values of $n4$ and $n5$ resemble the bubbles in the input images.
Notably, the distributions of $n4$ and $n5$ are opposite to one another, with the distributions of the positive values in $n4$ to those of the negative values in $n5$.
This is because the elements of $\bf{W}_\textrm{out}$ that correspond to $n4$ and $n5$, i.e., ${W}_{\textrm{out},4}$ and ${W}_{\textrm{out},5}$, have different signs, with ${W}_{\textrm{out},4}$ having a negative value when ${W}_{\textrm{out},4}$ has a positive value.
Moreover, the distribution in $n4$ is more obvious than that in $n5$. 
Thus, the discussion below is focused on $n4$.

In the image of the weight matrix for $n4$ shown in Fig.~\ref{fig:qout_s1}, positive elements indicate
``Fast'' features and negative elements indicate ``Slow'' features.
This is explained by analyzing the calculation process as follows. 

As shown in Fig.~\ref{fig:FNN}, there are two neurons in the output layer. 
We express values for ``Slow'' using the superscript ``s'' and ``Fast'' using the superscript ``f'', i.e., $\mathbf{y_\mathrm{pred}}=y_\mathrm{pred}^\mathrm{s}$ for ``Slow'' and $\mathbf{y_\mathrm{pred}}=y_\mathrm{pred}^\mathrm{f}$ for ``Fast'' neurons.

When ``Slow'' is predicted, for the output layer, $y_\mathrm{pred}^\mathrm{s}$ is larger than $y_\mathrm{pred}^\mathrm{f}$, i.e., $y_\mathrm{pred}^\mathrm{s}$ is closer to 1.
Therefore, $q_\mathrm{out}^\mathrm{s}$ has to be larger than $q_\mathrm{out}^\mathrm{f}$.
From Eq.~\ref{eq:s1toqout}, $q_\mathrm{out}^\mathrm{s}$ can be expressed as
\begin{equation} 
\label{eq:s1toqoutapp}
q_\mathrm{out}^\mathrm{s} = W_\mathrm{out,1}^\mathrm{s}s_\mathrm{1,1}+W_\mathrm{out,2}^\mathrm{s}s_\mathrm{1,2}+\dots+W_\mathrm{out,5}^\mathrm{s}s_\mathrm{1,5}+b_\mathrm{out}^\mathrm{s}.
\end{equation}
\noindent In Eq.~\ref{eq:s1toqoutapp}, since the $W_\mathrm{out,4}^\mathrm{s}$ obtained from the training process has a positive value, $s_\mathrm{1,4}$ has to be as large as possible to obtain the ``Slow'' prediction.
Moreover, the relationship between $s_\mathrm{1,4}$ and $q_\mathrm{1,4}$ is determined by the Leaky ReLU function as shown in Eq.~\ref{eq:hidden}.
Therefore, $q_\mathrm{1,4}$ has to be as large as possible.
From Eq.~\ref{eq:xtoq1}, $q_\mathrm{1,4}$ is expressed as
\begin{equation} 
\label{eq:xtoq1app}
q_\mathrm{1,4}=\mathbf{W_\mathrm{1,4}x}+b_\mathrm{1,4},
\end{equation}
where $\mathbf{W_\mathrm{1,4}}$ is the fourth row of the weight matrix $\mathbf{W_\mathrm{1}}$ and $b_\mathrm{1,4}$ is the corresponding bias element, which has an insignificant value in our case.
From above, $\mathbf{W_\mathrm{1,4}x}$ has to be as large as possible when ``Slow'' is predicted.

Since shadowgraphy was used to capture the image sequences of the laser-induced microjet, the pixels in an image covered by a vapor bubble or cavitation bubbles are dark, and thus have low normalized intensity values ($x_{i,j}$, the elements of $\mathbf{x}$, are close to 0).
On the other hand, due to the backlight, the pixels not covered by bubbles are bright and have high normalized intensity values ($x_{i,j}$ are close to 1).
Therefore, the elements of $\mathbf{W_\mathrm{1,4}}$ that correspond to the area covered by the bubbles in the ``Slow'' images but not that in the ``Fast'' images have negative values, so these values become zero when multiplied by a ``Slow'' image to increase the value of $\mathbf{W_\mathrm{1,4}x}$.
In contrast, for a ``Fast'' image, these values would remain as negative values, thereby reducing the value of $\mathbf{W_\mathrm{1,4}x}$.
Therefore, the elements with negative values, which are indicated by the blue pixels in Fig.~\ref{fig:qout_s1}, show the positions of the bubbles in the ``Slow'' images.
On the other hand, the elements of $\mathbf{W_\mathrm{1,4}}$ that correspond to the area covered by the bubbles in the ``Fast'' images but not in the ``Slow'' images have to be positive values so that these values would become zero when multiplied by a ``Fast'' image to reduce the value of $\mathbf{W_\mathrm{1,4}x}$.
However, for a ``Slow'' image, these values would remain as negative values, thereby increasing the value of $\mathbf{W_\mathrm{1,4}x}$.
These elements with positive values, which are indicated by the red pixels in Fig.~\ref{fig:qout_s1}, show the positions of the bubbles in the ``Fast'' images.

\section{Results and discussion}\label{sec:RanD}
In this section, we demonstrate the relationship between the jet velocity obtained from the experiments and that predicted by the FNN in Sec.~\ref{subsec:qout_result_main}. 
Next, in Sec.~\ref{subsec:qout_cavvap}, we explain the influence of a vapor bubble and cavitation bubbles on the jet velocity.
Finally, we report the extracted features related to the distribution of the cavitation positions obtained from the weight visualization in Sec.~\ref{subsec:W_vis}.

\subsection{The relationship between the experimental and predicted jet velocity}\label{subsec:qout_result_main}

In this subsection, we investigate and discuss the relationship between the jet velocity and the predicted values $\mathbf{q_\mathrm{out}}$ obtained from the FNN.

\begin{figure}[t]
\centering
\includegraphics[width=1.0\columnwidth]{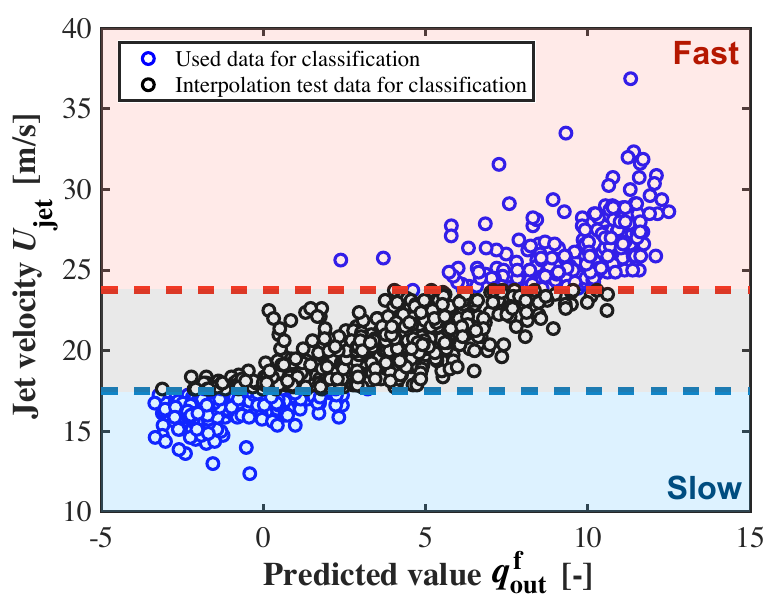}
\caption{\label{fig:Uvsqout}
The experimental jet velocity $U_\mathrm{\mathrm{jet}}$ and the predicted values $q_\mathrm{out}^\mathrm{f}$ for the FNN. The blue and red areas represent the data of the fastest 25\% (``Fast'') and the slowest 25\% (``Slow'') of the jet velocity, respectively.}
\end{figure}

As described in Sec.~\ref{subsec:FNN}, the classification of ``Fast'' and ``Slow'' is based on the vector $\mathbf{y_\mathrm{pred}}$, which is normalized by the sigmoid function, as shown in Eq.~\ref{eq:output}.
The sigmoid function saturates extreme negative and positive values to zero and one, respectively.
Therefore, we used the vector of $\mathbf{q_\mathrm{out}}$, which is before normalization, in the investigation of the relationship between the experimental values and the predicted values.
The relationship between the experimental jet velocity $U_\mathrm{\mathrm{jet}}$ and the $q_\mathrm{out}^\mathrm{f}$ is shown in Fig.~\ref{fig:Uvsqout}.
Note that $\mathbf{q_\mathrm{out}}$ has two values, but since both of them show similar trends, we use the values that indicate ``Fast''.

We also tested data other than the ``Fast'' and ``Slow'' data of the model used in this study as interpolation test data and obtained $q_\mathrm{out}^\mathrm{f}$ for each data point. Then, the relationship between the jet velocity and the predicted values for the interpolation test data was plotted as a black plot in Fig.~\ref{fig:Uvsqout}.
Remarkably, the correlation coefficient is 0.912, which is significantly higher than that obtained using the total area of the cavitation bubbles shown in Fig.~\ref{fig:exp_result}(b).
This result indicates that the FNN can predict the jet velocity more accurately, even for interpolation test data, by considering not only the total area of cavitation bubbles but also other features.
Therefore, we hereafter extract the features that the FNN observes to understand the important factors that influence the jet velocity.

\subsection{The influence of vapor bubbles vs. cavitation}\label{subsec:qout_cavvap}

\begin{figure}[t]
\centering
\includegraphics[width=0.9\columnwidth]{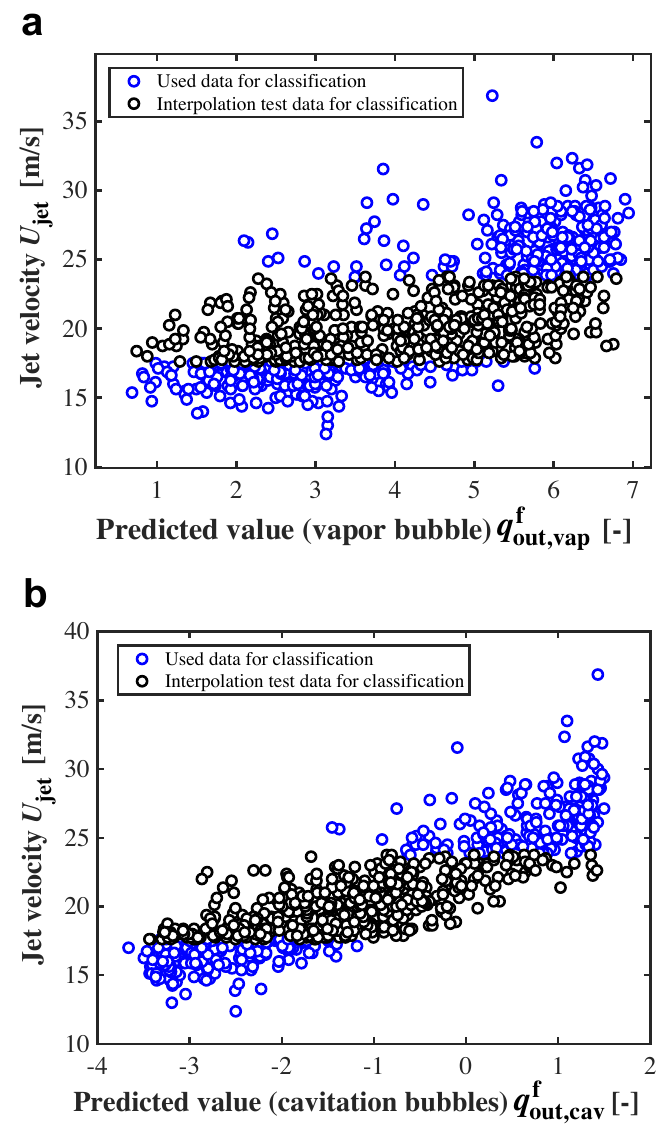}
\caption{\label{fig:Uvsqout_cavvap}
(a) The experimental jet velocity $U_\mathrm{\mathrm{jet}}$ vs. predicted values $q_\mathrm{out,vap}^\mathrm{f}$ for the FNN trained with images of vapor bubbles and (b) the experimental jet velocity $U_\mathrm{jet}$ vs. predicted values $q_\mathrm{out,cav}^\mathrm{f}$ for the FNN trained with images of cavitation bubbles.}
\end{figure}

In this subsection, we compare the effects of a vapor bubble and cavitation bubbles on the jet velocity by comparing the relationship between the experimental jet velocity and the predicted values from the FNN.
We compare the training of the FNN using cropped images of only a vapor bubble and only cavitation bubbles, respectively. 
Fig.~\ref{fig:Uvsqout_cavvap}(a) shows the relationship between the jet velocity $U_\mathrm{\mathrm{jet}}$ and the predicted values $q_\mathrm{out,vap}^\mathrm{f}$ when training only with a vapor bubble.
In addition, Fig.~\ref{fig:Uvsqout_cavvap}(b) shows the relationship between the jet velocity $U_\mathrm{\mathrm{jet}}$ and the predicted values $q_\mathrm{out,cav}^\mathrm{f}$ when training only with cavitation bubbles. 
The correlation coefficients are 0.688 and 0.883, respectively, and the averaged total accuracies of the classification are 90.0\% and 98.4\%, respectively. 
These results suggest that cavitation bubbles at $t=20$ \textmu s have a greater impact on the prediction of the jet velocity than vapor bubbles.

\subsection{Weight visualization: distribution of the cavitation position}\label{subsec:W_vis}

\begin{figure}[t]
\centering
\includegraphics[width=0.9\columnwidth]{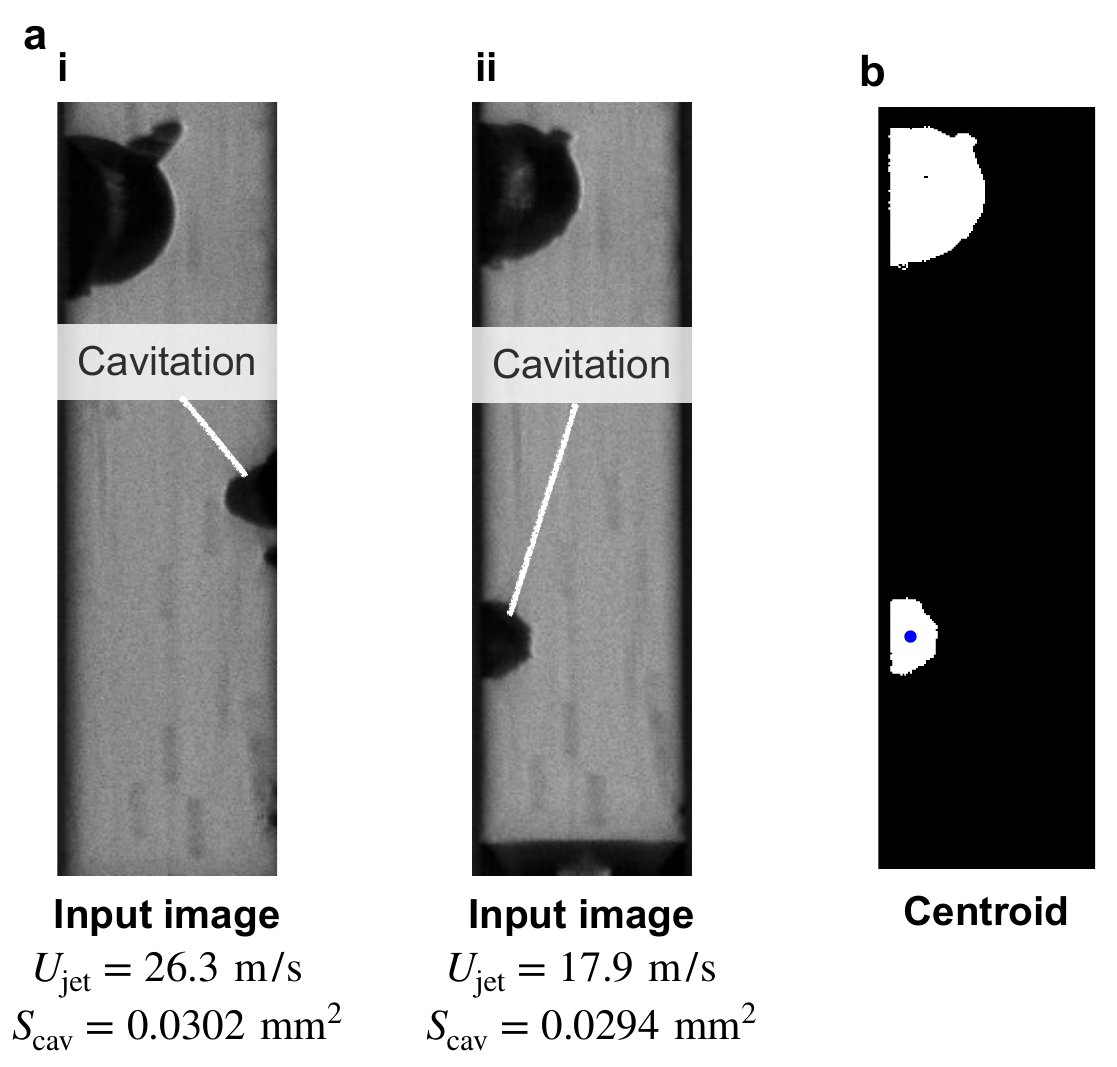}
\caption{\label{fig:input_cog}
(a) Input images of cases with similar sizes of cavitation bubbles occurring at different positions: (i) $S_\mathrm{cav}=0.0302$ mm$^2$, $U_\mathrm{jet}=26.3$ m/s, (ii) $S_\mathrm{cav}=0.0294$ mm$^2$, $U_\mathrm{jet}=17.9$ m/s and (b) the binarized image of (a)(ii) with the blue dot for centroid of the cavitation bubbles.}
\end{figure}

\begin{figure}[t]
\centering
\includegraphics[width=0.8\columnwidth]{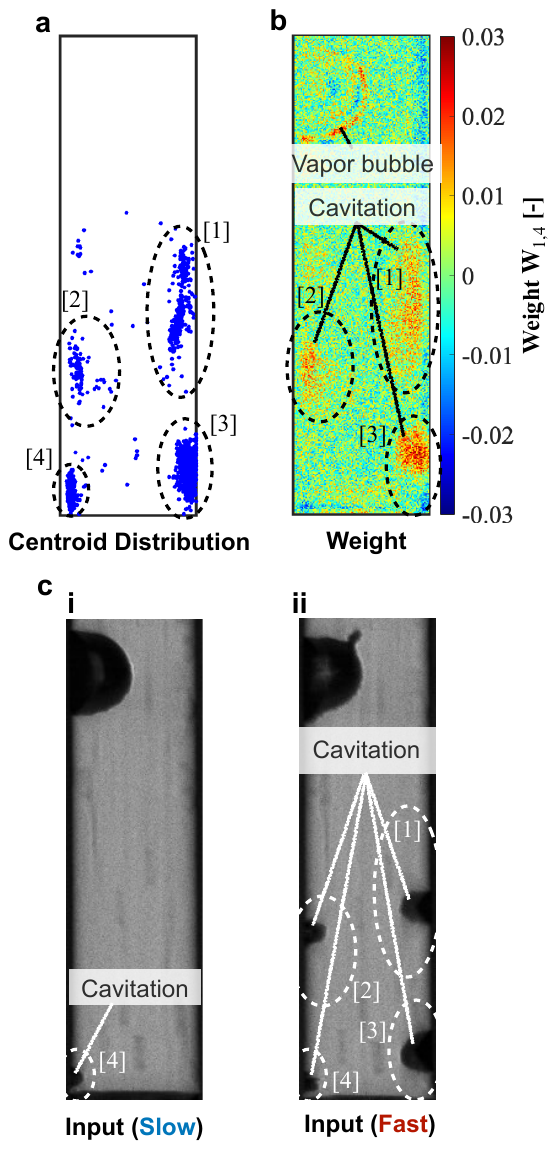}
\caption{\label{fig:weight_cog}
(a) The distribution of the centroids of the cavitation bubbles, (b) the colormap of the weight matrix $\mathbf{W_\mathrm{1,4}}$ for classification of the jet velocity, and (c) an example image of (i) ``Slow'' and (ii) ``Fast'' images, which include cavitation bubbles.}
\end{figure}

In this subsection, we describe an investigation of the feature extraction results and the effects of the cavitation position on the jet velocity.
For this purpose, we investigate the weight matrix $\mathbf{W_\mathrm{1}}$, which determines the predicted values $\mathbf{q_\mathrm{out}}$.

To demonstrate the effects of the cavitation positions, we compare cases with cavitation bubbles of similar sizes occurring at different places in Figs.~\ref{fig:input_cog}(a): (i) $S_\mathrm{cav}=0.0302$ mm$^2$ and (ii) $S_\mathrm{cav}=0.0294$ mm$^2$.
Remarkably, the jet velocity for the case of Fig.~\ref{fig:input_cog}(a)(i) is $U_\mathrm{jet}=26.3$ m/s ($q_\mathrm{out,cav}^\mathrm{f} = 0.332$), which is higher than the value of $U_\mathrm{jet}=17.9$ m/s ($q_\mathrm{out,cav}^\mathrm{f} = -1.70$) in Fig.~\ref{fig:input_cog}(a)(ii).
These cases show that the jet velocity greatly depends on the cavitation positions.
To further investigate the effects of the cavitation positions, the centroids of the cavitation bubbles are obtained from the binarized image as shown in Fig.~\ref{fig:input_cog}(b).
The centroids of the cavitation bubbles for all data (1000 data) are shown in Fig.~\ref{fig:weight_cog}(a).
From this figure, we identified regions where the cavitation is often located.
Regions where cavitation bubbles are frequently generated are marked as Regions [1], [2], [3], and [4] in the figure.
Out of the 1000 data, cavitation bubbles are observed in 931 (93.1\%) data, including 250 out of the 250 (100\%) ``Fast'' data, 188 out of the 250 (75.2\%) ``Slow'' data, and 493 out of the 500 (98.6\%) data that are neither ``Fast'' nor ``Slow''.
For comparison with the extracted features, Fig.~\ref{fig:weight_cog}(b) shows an image of the weight matrix $\mathbf{W_\mathrm{1,4}}$ obtained using the method described in Sec.~\ref{subsec:vis_pro}.
Note that positive weights (red region) indicate ``Fast'' features and negative weights (blue region) indicate ``Slow'' features, as explained in Sec.~\ref{subsec:vis_pro}.
Comparing Figs.~\ref{fig:weight_cog}(a) and (b), we verify the ``Fast'' weight in Regions [1]--[3].
Note that the ``Fast'' weight is also shown around the vapor bubble. 
This demonstrates that the larger the vapor bubble, the faster the jet velocity, which is consistent with a finding of the previous study by Kawamoto \textit{et al.}\cite{Kawamoto2016volume}

Although the ``Fast'' weight can be seen in Regions [1]--[3], 
``Fast'' features are not seen in Region [4] in Fig.~\ref{fig:weight_cog}(b).
To investigate this region, we show an example of a ``Slow'' input image, although the cavitation bubbles are generated in Region [4] in Fig.~\ref{fig:weight_cog}(c)(i).
In fact, cavitation bubbles occurred in Region [4] in 152 out of the 250 (60.8\%) ``Slow'' images.
This shows that there are cavitation positions that are not related to the increase in the jet velocity.
Note that, as can be seen from the example of a ``Fast'' input image indicating the generation of cavitation bubbles at all defined regions shown in Fig.~\ref{fig:weight_cog}(c)(ii), we can also observe the bubble generation at Region [4] even in the ``Fast'' image.
Therefore, Figs.~\ref{fig:weight_cog}(c)(i) and (ii) indicate that the bubble generations in Regions [1]--[3] affect the increase in jet velocity, while the generation at Region [4] is hardly related to the jet velocity.

\begin{figure}[t]
\centering
\includegraphics[width=1.0\columnwidth]{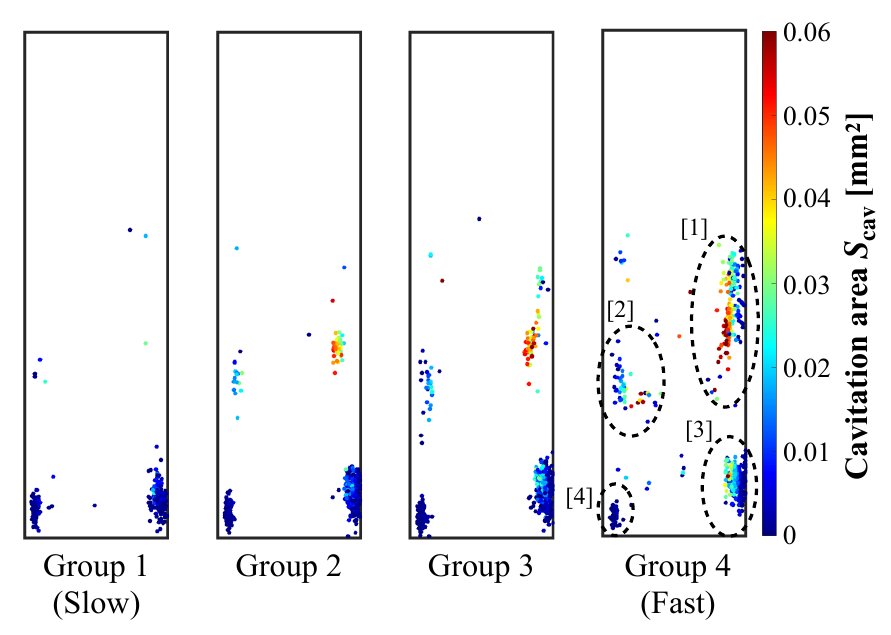}
\caption{\label{fig:cog_pos}
The distribution of the centroids of the cavitation bubbles with a colormap of the cavitation area $\mathbf{S_\mathrm{cav}}$ for 250 data at each velocity. Groups 1,2,3 and 4 are in order of decreasing velocity.}
\end{figure}

Here, the weight diagram shown in Fig.~\ref{fig:weight_cog}(b) indicates the overall evaluation of the frequency of cavitation generation in each region and the cavitation area.
The more frequent the cavitation generation in each region, the greater the value of the weight element at that position.
Moreover, the larger the cavitation area, the higher the possibility that the weight of the region is large.
In order to distinguish between the cavitation frequency and the cavitation area at each position, the area of each cavitation is represented by the scale of the colorbar and plotted at the centroid in Fig.~\ref{fig:cog_pos}.
As shown in the subfigures, we divide the data into 4 groups, each 250 data, in order of ascending velocity.
The slowest 25\% are Group 1 (``Slow''), the next slowest 25\% are Group 2, the second fastest 25\% are Group 3, and the fastest 25\% are Group 4 (``Fast''). 
Fig.~\ref{fig:cog_pos} shows that the cavitation area increases with the jet velocity.
This is consistent with what was described in the previous study of Kiyama \textit{et al.} \cite{kiyama2016effects}

Fig.~\ref{fig:cog_pos} also shows that the higher the jet velocity, the more cavitations occur in Regions [1] and [2].
We show the number of cavitation generations in each region and the velocity range in Table~\ref{tab:cav_num}.
In Regions [1] and [2], the number of cavitation bubbles increases with increasing velocity.
On the other hand, the number does not necessarily increase in Regions [3] and [4].
This suggests that the effect of the velocity increase is more significant when cavitation occurs closer to the laser focus position, where it is farther from the gas-liquid interface.
Note that, in Region [3], cavitation bubbles with a large area are generated at high velocities, while in Region [4], small cavitation bubbles are generated regardless of the velocity range, so the weight is red in Region [3], and there is almost no weight in Region [4], as shown in Fig.~\ref{fig:weight_cog}(b).

Here, we discuss the possible mechanisms of how the cavitation position affects the jet acceleration.
Kiyama \textit{et al.}\cite{kiyama2016effects} reported that after the propagation of the compression wave, the fluid pressure becomes positive, while after the propagation of the expansion wave, it becomes negative.
Moreover, the compression waves are converted into expansion waves and vice versa at the gas-liquid interface.
When the waves are in close proximity to the gas-liquid interface, the time interval between the propagation of compression and expansion waves is minimal.
In contrast, the farther away it is from the gas-liquid interface, the longer the time interval is.
Therefore, the greater the distance from the gas-liquid interface, the longer the time the fluid experiences the tensile stress, which means a greater pressure impulse.
As reported by Hayasaka $et\ al$.\cite{hayasaka2017}, the larger the pressure impulse experienced by the liquid, the larger the jet velocity.
Consequently, the cavitation positions farther from the gas-liquid interface, where the fluid experiences the larger pressure impulse, makes a more significant contribution to the velocity increase.
For further understanding, we need to observe the propagation behavior of a shock wave with cavitation bubbles.
For this purpose, we suggest the use of the background-oriented schlieren (BOS) technique \cite{venkatakrishnan2004density, yamamoto2022contactless} to visualize the shock waves in the microtube.

\begin{table}[t]
\centering
\caption{\label{tab:cav_num} 
The number of cavitation bubbles in each region and for each velocity range.}
\begin{tabular}{ccccccccc}
\hline\hline
 & Region 1 & Region 2 & Region 3 & Region 4 & Total\\
\hline
Group 1 & 3 (0.865\%) & 4 (1.15\%) & 177 (51.0\%) & 163 (47.0\%) & 347\\ 
Group 2 & 39 (8.52\%) & 17 (3.71\%) & 262 (57.2\%) & 140 (30.6\%) & 458\\
Group 3 & 76 (13.8\%) & 30 (5.44\%) & 307 (55.7\%) & 138 (25.0\%) & 551\\
Group 4 & 227 (36.7\%) & 50 (8.09\%) & 285 (46.1\%) & 56 (9.06\%) & 618\\
\hline\hline
\end{tabular}
\end{table}

\section{Conclusions and outlook}\label{sec:conclusion}

In this study, we have conducted 1,000 experiments to obtain corresponding images and jet velocities.
We have investigated the relationship between the amount of cavitation bubbles and the jet velocity and found that there was still non-negligible variation.

To investigate the reason for this, we utilized explainable artificial intelligence (XAI), namely a feedforward neural network (FNN), to make predictions and classify the jet velocity from images of cavitation bubbles.
The results showed that at $t=20$ \textmu s, when cavitation was most likely to occur, the averaged total accuracy was as high as 99.2\%.
This confirmed that the FNN extracted clear features related to the jet velocity.

In order to clarify the extracted features, we visualized the calculation process in the FNN.
The results suggested that the prediction values obtained by training only cavitation bubbles had a higher correlation with the jet velocity than those using only vapor bubbles.
This demonstrates that cavitation bubbles have a greater impact on predicting the jet velocity than vapor bubbles.
In addition, we verified that the correlation coefficient between the experimental jet velocity and the predicted values obtained from training only cavitation bubbles was higher than that between the jet velocity and the area of the cavitation bubbles.
This indicates that the FNN extracted features other than the area of cavitation.

Therefore, we visualized the ``weights'' that represent the important signals in the input images to investigate the extracted features.
As a result, we clarified that there are cavitation positions that do or do not affect the jet velocity.
Further investigation suggested that cavitation that occurs closer to the laser focus position has a higher acceleration effect.
We argue that this is because in close proximity to the laser focus position, the greater the pressure impulse.

We now summarize three possible contributions that could be developed from the findings of this study.
The first is the development of the jet velocity prediction method.
We successfully predicted the jet velocity from the images of jet generation using machine learning, indicating that machine learning can be used as an application for predicting jet velocities.
The second contribution is the development of XAI used in this study.
We successfully extracted features from a large amount of data acquired from experiments, verifying the usefulness of XAI as an aid in the interpretation of fluid experiment images.
We believe that XAI is a highly applicable method for visualizing important features, especially in further experimental fluid systems with many parameters. 
Finally, the third contribution is the usefulness of this approach not only for high-speed microjets but also for various phenomena involving cavitation bubbles.
For this, we need to further consider the physical meaning of the critical cavitation position.
Cavitation bubbles and the jet velocity are related to the propagation of shock waves.
Therefore, we need to observe the propagation behavior of a shock wave with cavitation bubbles.
For this purpose, a possible method to visualize the shock waves in a microtube is the background-oriented schlieren (BOS) technique\cite{venkatakrishnan2004density, yamamoto2022contactless}.

\begin{acknowledgments}

This work was funded by the Japan Society for the Promotion of Science (Grant Nos. 20H00223, 20H00222, and 20K20972) and the Japan Science and Technology Agency PRESTO (Grant No. JPMJPR21O5). The authors would also like to thank Dr. Masaharu Kameda (Professor, Tokyo University of Agriculture and Technology) and Dr. Akinori Yamanaka (Professor, Tokyo University of Agriculture and Technology) for helpful discussions and comments.

\end{acknowledgments}

\section*{References}

\nocite{*}
\bibliography{aipsamp}

\end{document}